# A computational study on imputation methods for missing environmental data

P. Dixneuf[1], F. Errico[1,2], M. Glaus[1]

[1]Station Expérimentale des Procédés Pilotes en Environnement, École de Technologie Supérieure, Université du Québec, 1100 rue Notre-Dame Ouest, Montréal, QC H3C 1K3, Canada
[2]Groupe d'études et de recherche en analyse des décisions , Pavillon André-Aisenstadt, Montréal, QC H3T 1J4, Canada

**ABSTRACT.** Data acquisition and recording in the form of databases are routine operations. The process of collecting data, however, may experience irregularities, resulting in databases with missing data. Missing entries might alter analysis efficiency and, consequently, the associated decision-making process. This paper focuses on databases collecting information related to the natural environment. Given the broad spectrum of recorded activities, these databases typically are of mixed nature. It is therefore relevant to evaluate the performance of missing data processing methods considering this characteristic. In this paper we investigate the performances of several missing data imputation methods and their application to the problem of missing data in environment. A computational study was performed to compare the method missForest (MF) with two other imputation methods, namely Multivariate Imputation by Chained Equations (MICE) and K-Nearest Neighbors (KNN). Tests were made on 10 pretreated datasets of various types. Results revealed that MF generally outperformed MICE and KNN in terms of imputation errors, with a more pronounced performance gap for mixed typed databases where MF reduced the imputation error up to 150%, when compared to the other methods. KNN was usually the fastest method.  MF was then successfully applied to a case study on Quebec wastewater treatment plants performance monitoring. We believe that the present study demonstrates the pertinence of using MF as imputation method when dealing with missing environmental data.

*Keywords*: computational study, environment, missing data, missForest, imputation

## 1. Introduction

Recent developments in data science and analytics show that decision-making processes can greatly take advantage of the availability of large amount of well-organized data in the form of databases. Databases are analyzed to extract relevant knowledge. However, because failures may occur in the acquisition and/or recording processes, some of the data in the databases can be missing. Missing data can arise from numerous reasons (equipment failure, human operator error, manual data entry mistakes, etc.) and researchers and practitioners of many fields (medicine, industrial production, education, environment, etc.) are faced with this issue. For instance, in the medical literature approximatively 89% of all clinical experiments have to cope with incomplete datasets (Wood et al., 2004). Missing data may induce a loss in statistical power of the database analysis (Waljee et al., 2013), which can bias the subsequent decision-making. Moreover, missing data impedes the adoption of established methods of analysis that require a complete dataset (Liao et al., 2014). Because recollecting data is often not an available option due to physical or financial constraints, a substantial amount of research has recently been devoted to this issue.

Amongst the different existing missing data processing methods, the most conventional ones opt for the deletion of records lacking one or several attributes. Because case deletion methods might result in the deletion of relevant attributes, these methods are, however, generally not recommended (Sessa and Syed, 2017). Consequently, data scientists turned their attention toward the development of so-called imputation methods (IMs). Imputation is the process of substituting missing data by guessed values, called "imputed data". There are several methods to accomplish this, some based on basic mathematical principles, like the mean imputation method, while others rely on more advanced statistical approaches and machine learning. A given IM may be more suitable for a given dataset than another, depending on how the imputation is performed and the application context. This study is interested in investigating several IMs and their performances when applied to environmental contexts.

Organizing environmental data in well-structured databases is a challenging task (Blair et al., 2019). On the one hand, the natural environment is impacted by human activities, and this calls for interdisciplinary research and analysis. On the other hand, natural phenomena cover different time and spatial scales and are generally interconnected, which makes data integration difficult. This typically results in heterogeneous data sources and generally gives rise to databases of a mixed nature, with both qualitative and quantitative entries. Because of these specificities, the treatment and analysis of data related to environmental systems represents nowadays a real challenge. This is especially true if we consider that in practice the personnel called to treat such data are not specialized data scientists. We will see later how the above considerations guided several choices in the methodology followed in the present work, including the choice of the IMs on which to focus, the preferred performance measures, and the experimental setup.

The scientific literature on IMs and their performance is abundant, however a thorough search yielded no study addressing all the challenges associated with environmental contexts. Depending on the objectives and target discipline of each publication, IMs were investigated under different conditions. For instance, some authors performed their study on one specific dataset (Noor et al., 2015; Roda et al., 2014), the provided analyses are therefore not easily extendable to other datasets. Other works considered one type of data only, as for example Wu et al. (2015), Schmitt et al. (2015), and Ghorbani and Desmarais (2017), who focused on qualitative, quantitative and binary data, respectively, while this is in contrast with the general mixed nature of environmental data sets. Furthermore, as mentioned in Solaro et al. (2014), the quality of IMs performances and their sensitivity to the characteristics of the databases, such as dimension, correlations, etc. has not been adequately investigated in the literature. Additionally, computational studies were generally carried out involving extensive calibration work upon the parameters of each IM (Gromski et al., 2014; Stekhoven and Bühlmann, 2012; Wu et al., 2015), thus requiring advanced skills in programming for implementation.

In this paper, to assess the applicability of IMs to environmental contexts, especially in view of heterogeneous sets of databases, an experimental campaign comprised of two main phases was performed. In the first one, we selected 10 datasets from various sources in the literature and artificially obtained various degrees of missing data by randomly removing some of the entries. The set of selected databases was chosen to be representative of the typical characteristics found when analysing environmental data, such as varying dimensions, as well as heterogeneous data types and structural features. We then selected three well-known IMs for comparison, all sharing the ability to treat databases with mixed data types, namely missForest (MF) (Stekhoven and Bühlmann, 2012), Multivariate Imputation by Chained Equations (MICE) (Buuren and Oudshoorn, 1999) and K-Nearest Neighbors (KNN) (Troyanskaya et al., 2001). We tested these IMs on the obtained incomplete datasets with missing data and compared their performances in terms of quality of imputation, as well as computing times. Relations between the structural characteristics of the datasets and the performance of the IMs were also investigated. One of the major insights gained in this part of the study is the superiority of imputation error obtained with MF when compared with MICE and KNN. It is worth noticing that we purposely chose to perform no ad hoc parameter calibration when comparing the selected IMs.

In the second phase of the experimental campaign, a case study was performed on the database collecting Quebec's wastewater treatment plants performance made available by the Ministère de l'Environnement et de la Lutte contre les Changements Climatiques (MELCC). Its accuracy and veracity are guaranteed until 2013 for all the 814 treatment plants included in the database. However, because the measurement frequency of these parameters varied from station to station, the database is incomplete. Being able to impute the missing data would allow governmental decision makers to perform an exhaustive interannual analysis of the Quebec wastewater treatment plants based on the monitored parameters, and consequently, a to gain a better understanding of one of Quebec's most valuable natural resources. Coherently with the insights gained in the first experimental phase, the case study was performed by adopting the best performing IM, namely MF.

The rest of the document is organised as follows. Section 2 presents the performance comparison of KNN, MICE and MF. It describes the studied IMs, the datasets, and the process that was used to compare them before giving the results of the comparison. Section 3 presents the case study. It contains a description of the Quebec wastewater treatment plants and introduces the indicator we used to assess imputation accuracy before giving the results of the imputation. In Section 4 we draw our conclusions.

## 2. Comparative study of imputation methods

In order to evaluate and compare the performances of MF, MICE and KNN and their applicability to environmental contexts, a comparative study was performed. The comparison process is presented in this section into five parts. Section 2.1 goes into the details about the selected IMs and the classification scheme used to differentiate them. In Section 2.2, the datasets on which the IMs were compared are briefly described. In Section 2.3, the approach to assessing the IMs performance is explained. Finally, the results of the performance comparison are given in Section 2.4 and discussed in Section 2.5.

### 2.1. Imputation methods

2.1.1. Categories of imputation methods

In missing data treatment literature, two schemes of classification are broadly used for IMs (Little and Rubin, 2002): (1) single and multiple imputation methods and (2) parametric and non-parametric methods. These categories are described below.

With single imputation methods, a missing entry is imputed one single time, meaning that the only imputed data is treated as the true data that would be observed if the dataset would have been complete, which is not necessarily true. Consequently, single IMs do not consider the uncertainty of imputed data (Zhang, 2016). On the opposite, multiple imputation methods repeat the imputation process several times, resulting in multiple imputed datasets (Little, RJA and Rublin, 1987). Therefore, to each missing entry is associated multiple imputed data that are all likely to be the desired outcome. These multiple datasets are then analyzed and combined into a single imputed dataset. Because the statistical analysis calculates the variation in parameter estimates between the different imputed datasets, multiple imputation methods incorporate the uncertainty inherent in the imputation.

Parametric IMs impute missing data by making assumptions on data distribution based on the observed data, thus assuming the complete data can be modeled by a probability distribution with a fixed set of parameters. This approach can induce a bias because data distribution assumptions may not be verified. For instance, some parametric methods such as linear regression define quantitative attributes only through linear terms with no interactions, but relevant non-linear terms may be omitted, resulting in biased results (Seaman et al., 2012). Furthermore, due to collinearity, parametric methods may demonstrate difficulties imputing a dataset in which two variables are highly correlated (Shah et al., 2014). On the other hand, non-parametric IMs do not rely on distribution assumptions, meaning that the parameters of the studied population are no longer fixed.

2.1.2. Selected imputation methods

MF, MICE and KNN were selected because they represent various forms of IMs categories and are widely adopted in the scientific literature. The key feature they have in common is the ability to impute qualitative and/or quantitative datasets. The main characteristics of the selected IMs are summarized in Table 1.

**Table 1.** Main characteristics of the selected IMs

| Imputation methods | Parametric | Multiple Imputation |
|---|---|---|
| KNN | No | No |
| MICE | Yes | Yes |
| MF | No | No |

KNN is an IM based on the nearest neighbor search that was initially introduced for the study of gene expression (Troyanskaya et al., 2001). The imputation of an observation that lacks one or several attributes starts by finding the $k$ nearest neighbors of the said observation. The search process is done by computing the distance between each pair of observations that contains the observation of interest, with the distance being based on the complete attributes. The data is then imputed based on the $k$ nearest neighbors that possess the specific attribute. For a classification, a majority vote is applied on the attributes of the nearest neighbors, and a mean in the case of a regression. In order to take into account both qualitative and quantitative variables, the Gower distance is applied between the pairs of observations (Gower, 1971).

MICE is a multiple imputation method introduced by Van Buuren (1999). It uses an iterative procedure where each variable is imputed on the other variables. At each iteration, the imputed data is updated and so are the predictors, and the imputation becomes more and more precise. Because it uses the multivariate imputation scheme, this iterative procedure is repeated $m$ times, resulting in $m$ imputed datasets before aggregating them into one. One of the particularities of MICE is the possibility of using a specific imputation model for each imputed variable, depending on its type and on the user's choice. By default, a predictive mean matching is used for quantitative data, a logistic regression imputation for binary data, a polytomous regression imputation for unordered qualitative data and a proportional odds model for ordered qualitative data. Because of the imputation models used by MICE, it is a parametric IM.

MF is a non-parametric IM based on random forests (Stekhoven and Bühlmann, 2012). Random forests (Breiman, 2001) are predictive models combining two main ideas: decision trees and bootstrap aggregating (bagging). A decision (classification or regression) tree consists in a series of tests on the observed data of a population to draw a conclusion about a targeted missing entry. Their goal is to underline generalizable data patterns that can be used to make inferences. Although a decision tree is usually built on the entire observable dataset, bagging suggests building $t$ decision trees on $t$ random subsets. The targeted result is then a combination of the results given by all the trees: a majority vote for classification and a mean for regression. Because aggregating trees reduces the influence of entries that could induce a bias, it improves model stability and precision (Breiman, 1996a; Domingos, 1997; Grandvalet, 2004). To impute an incomplete dataset, MF builds a random forest on each variable using the rest of the variables as predictors. This process is then iteratively repeated with the imputed dataset being updated at each iteration until a stopping criterion is met. Furthermore, through bagging, MF allows for an Out-Of-Bag (OOB) estimate of the imputation error called OOB error (Breiman, 1996b). This estimate is computed from the ability of the model to recover withdrawn observed data. The OOB procedure can be divided in 4 steps: (i) an entry is taken out of the observed data and is therefore set as "missing"; (ii) the withdrawn entry is imputed by using the decision trees which were not built using it; (iii) the two previous steps are repeated for all the observed data; (iv) the imputation error is then estimated as the mean of the differences between the observed and recovered data.

### 2.2. Datasets

As previously mentioned, given that the natural environmental is impacted by a large number of factors including natural, as well human activities, it is not a surprise that databases in this discipline considerably vary in terms of dimension (number of rows and columns), attributes type and structural features. In this spirit, the datasets used in the comparative study have been chosen in a way they reflect this diversity. Table 2 enumerates and briefly describes the selected databases, all coming from well-known sources in literature. Two of these datasets are present on the basic version of R, an environment for statistical computing and graphics (R Development Core Team, 2016), namely, the Iris dataset, which has been used several times in this kind of study (Aljuaid and Sasi, 2017; Misztal, 2013; Stekhoven and Bühlmann, 2012) and the Lanza and Rock datasets. Seven other datasets come from the UCI machine learning repository (Bache and Lichman, 2013), which provides access to 440 real datasets.

**Table 2.** Characteristics of the datasets used in the comparative study

| Datasets | Number of records (rows) | Number of attributes (columns) | Attributes type | Source | Abbreviation |
|---|---|---|---|---|---|
| Lanza | 100 | 3 | qualitative | R | Lanza |
| Hayes-Roth | 132 | 5 | qualitative | UCI | Hayes |
| Tic-Tac-Toe Endgame | 958 | 10 | qualitative | UCI | Tic-Tac-Toe |
| Rock | 48 | 4 | quantitative | R | Rock |
| Concrete Slump Test | 103 | 10 | quantitative | UCI | Concrete |
| Wine Quality | 122 | 12 | quantitative | UCI | Wine |
| Parkinsons | 195 | 22 | quantitative | UCI | Parkinson |
| Iris | 150 | 5 | mixed type | R | Iris |
| Contraceptive Method Choice | 313 | 10 | mixed type | UCI | Contraception |
| Musk | 476 | 167 | mixed type | UCI | Musk |

### 2.3. Approach to performance assessment

Figure 1 presents the principle of analysis that was followed to evaluate and compare the performances of the three studied IMs on the 10 complete datasets and for four different percentages of missing data (2, 5, 10 and 20%).

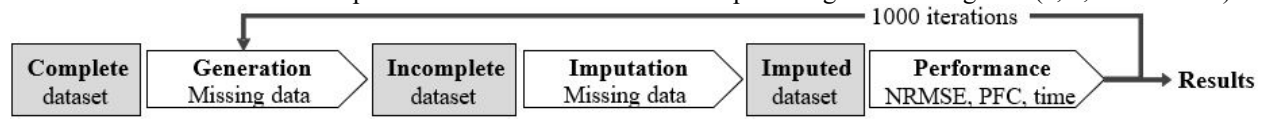

**Figure 1.** Performance assessment algorithm.

A percentage of missing data was artificially generated on the original datasets before imputing the resulting incomplete datasets with KNN, MICE and MF and assessing their performances. This missingness was generated completely at random with different missing data percentages. Simulations were carried out with missing data percentages up to 50% but were not included in the results because of a will for a synthesis (see Dixneuf, 2019, for details). The maximum threshold of 50% was applied based on preliminary results showing that for certain datasets, none of the IMs could impute the missing data. It can be observed that the generation of the incomplete dataset, as well as the imputation process are affected by randomness. To the scope of providing statistically accurate results, we made our analyses by repeating the whole experiment 1000 times and extracted suitable performance measures.

The behaviour of IMs may be regulated by a large set of parameters, such as the number of nearest neighbors for KNN, the number of multiple imputations for MICE, the number of trees grown in each random forest for MF, and many others. Researchers and data scientists may then fine tune and tailor the performance of a given IM to a specific context. However, this process typically requires the availability of specialized knowledge and computational infrastructure that are seldom accessible outside research laboratories. Furthermore, the obtained parameter settings are case specific. Given that our study aims at providing guidelines suitable for a general audience, including non-specialized users, and valid for a wide range of situations, we decided to avoid performing problem specific parameter optimization, and use instead the default values set in the publicly available programming software R.

2.3.1. Evaluation criteria

Following the imputation of the artificially incomplete datasets, the relative performance of each IM was assessed using two error indicators. The imputation errors associated with quantitative attributes was computed by the normalized root mean squared error – NRMSE (Oba et al., 2003). This indicator computes the difference between the real and the imputed dataset and can be superior to a 100%. The principle of normalization makes NRMSE suitable to compare datasets of varying sizes. Considering that the mean and variance are computed only on the missing data, NRMSE can be defined as follows:

$$NRMSE = \sqrt{\frac{mean\left[\left(X^{Complete} - X^{Imputed}\right)^2\right]}{var\left(X^{Complete}\right)}} \quad (1)$$

where $X^{Complete}$ is the complete dataset and $X^{Imputed}$ the imputed dataset. *Var* is used as short notation for empirical variance.

Concerning the imputation errors associated with qualitative attributes, the used indicator was the proportion of falsely classified entries – PFC, it can be defined as the number of falsely classified data divided by the total number of classified data. In the case of mixed type datasets, the performance of each IM was evaluated using these two indicators simultaneously.

2.3.2. Data structure descriptive criteria

Potential relationship between data structure and IMs performance was investigated. More specifically, the studied criteria were correlations of variables along with skew nature of distributions. Three moment-based indices (Solaro *et al.*, 2015) were used to describe the structural characteristics of the studied datasets. These indices were computed using the Bravais-Pearson linear correlation coefficient – $\rho_{XY}$ formulae:

$$\rho_{XY} = \frac{Cov(X,Y)}{\sigma_X \sigma_Y} \quad (2)$$

where X and Y are two independent variables. *Cov* is used as short notation for empirical covariance and $\sigma_X$ (resp. $\sigma_Y$) is the standard deviation of *X* (resp. *Y*). However, the Bravais-Pearson correlation coefficient can be applied only to quantitative attributes. The computation of the correlation coefficient between two attributes of different nature is based upon the conversion of qualitative attributes into quantitative ones (Zhang et al., 2015). This conversion assigns a numerical value to each class of the qualitative attribute. It is the mean value of the records in the quantitative attribute which belong to the corresponding class.

The Bravais-Pearson correlation coefficient gives knowledge about the strength of the relationship between two variables. It ranges from -1 to 1 and the threshold values (Wassertheil and Cohen, 1970) used in this study are given in Table 3.

**Table 3.** Bravais-Pearson correlation coefficient threshold values

| Absolute values of $\rho_{XY}$ | Strength of relationship |
|---|---|
| Around 0.8 | Strong |
| Around 0.5 | Moderate |
| Around 0.2 | Weak |

The first of the moment-based indices is the absolute mean correlation – $\rho_{abs}$. It is the absolute mean of all correlation coefficients in the datasets and gives knowledge about the overall magnitude of correlations between the variables of the datasets, regardless of the sign of the correlations. It can be defined as follows:

$$\rho_{abs} = \frac{2}{p(p-1)} \sum_{j=1}^{p-1} \sum_{l>j}^{p} |\rho_{jl}| \tag{3}$$

with $p$ the number of attributes in the dataset and $\rho_{jl}$ two variables of index by $j$ and $l$.

The second of the indices is the absolute standard deviation – $sd_{abs}$. It indicates if there is an unbalancing amongst the correlations and thus equals 0 when correlations are perfectly balanced. Its formulae is the following:

$$sd_{abs} = \sqrt{\frac{2}{p(p-1)} \sum_{j=1}^{p-1} \sum_{l>j}^{p} (|\rho_{jl}| - \rho_{abs})^2} \tag{4}$$

The last of the moment-based indices is the absolute skewness index – $skew_{abs}$. It provides information on the shape of the unbalancing between the correlations (positive or negative skewness). A positive absolute skewness index implies an unbalancing towards the left of the median and, conversely if the index is negative. It can be defined as:

$$skew_{abs} = \frac{\frac{2}{p(p-1)} \sum_{j=1}^{p-1} \sum_{l>j}^{p} (|\rho_{jl}| - \rho_{abs})^3}{sd_{abs}} \tag{5}$$

### 2.4. Results

The performance of MF, MICE and KNN were evaluated and compared on ten complete datasets, with a varying percentage of missing data (2, 5, 10, and 20%), considering three indicators: PFC (%), NRMSE (%) and processing times. Results are organized by data type (qualitative, quantitative, and mixed data type) and presented in box-and-whisker plot form. Each figure gives a statistical description of the results obtained over 1000 simulations, namely the lowest and highest data points, the first and third quartile and the median. Outliers were excluded for readability reasons.

2.4.1. Qualitative datasets

Figure 2 plots imputation errors (PFC) over 1000 imputations on the tree qualitative datasets for 2, 5, 10, and 20% missing data.

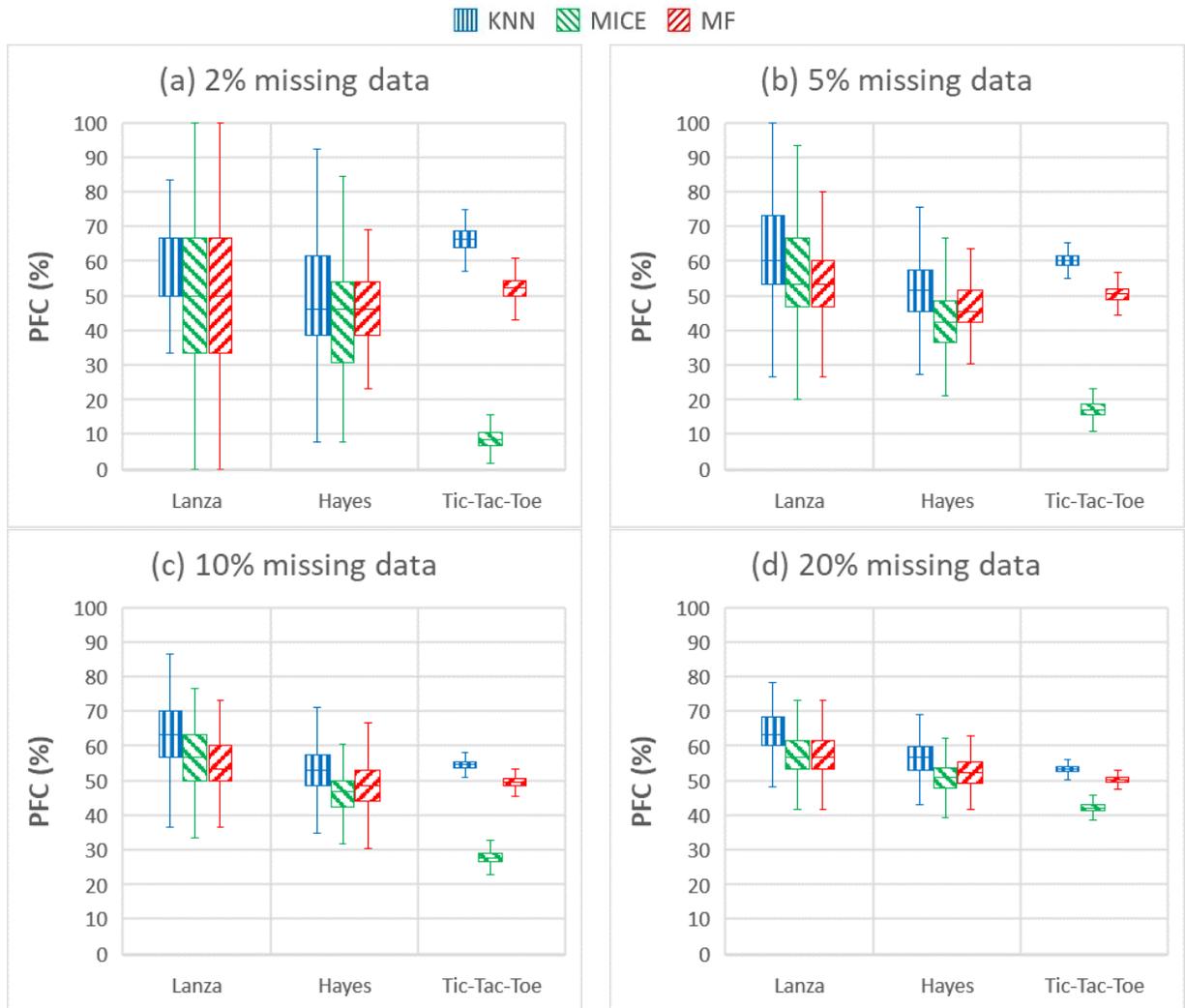

**Figure 2.** Boxplots of PFC (%) for the three IMs (KNN, MICE and MF) over the three qualitative datasets ("Lanza", "Hayes" and "Tic-Tac-Toe") for four different degrees of missing data: (a) 2%, (b) 5%, (c) 10% and (d) 20%.

The four graphs of Figure 2 show that the overall trend for imputation errors is to increase, on average, with increasing percentage of missing data. The one exception is the "Tic-Tac-Toe" case in which errors for KNN and MF are the highest when only 2% data are missing (on average respectively 66% and 52%) and while MF's performance remain relatively constant, KNN's errors decrease with increasing percentage of missing data, reaching around 53% imputation errors.

Observing the length of the boxes and whiskers of each graph, it appears that the standard deviation of imputation errors decreases with increasing percentage of missing data for the three IMs, in other words, the length of boxes and whiskers decreases. This may be explained by the fact that, the bigger the missing data percentage, the less probable it is for an IM to stumble upon a peculiar missingness layout that could produce especially low or especially high imputation errors, hence resulting in a scattered set of results. Additionally, all graphs show that, regardless of the IM or percentage of missing data, the length of the boxplots tends to reduce with increasing dataset size. This may be because the smaller the dataset, the less missing entries there are to impute, and consequently, the more probable it is to stumble upon entries that are especially hard or especially easy to impute.

Considering that the three qualitative datasets are arranged in ascending order of dimension (number records and attributes), imputation errors do not seem to be affected by the size of the imputed dataset. Indeed, apart from MICE, imputation errors were generally the lowest on the intermediate sized dataset, namely "Hayes". Although errors associated with the MICE method seem to decrease with increasing dataset dimension, this can be explained with "Tic-Tac-Toe" characteristics. In the case of the "Tic-Tac-Toe" dataset, imputation errors for MICE are lower at a low percentage of missing data regarding the other IMs but rise significantly from around 9% to 42% with increasing percentage of missing data. This can be explained by one of MICE's feature which is to specify a specific imputation model depending on the attribute's type. Here, in contrast with the two other datasets, the classification concerns a binary variable. The imputation model used by MICE is therefore different.

Even if KNN is systematically the least performing IM, neither MICE nor MF stands out from the other IMs. On average over the 1000 simulations, MF is the most performing IM on "Lanza" whereas MICE outperforms MF on "Hayes" and "Tic-Tac-Toe". However, because of the significant rise in MICE errors on the "Tic-Tac-Toe" case, it loses its advantage as the missing data percentage increases.

2.4.2. Quantitative datasets

Figure 3 plots imputation errors (NRMSE) over 1000 imputations on the four studied quantitative datasets for 2, 5, 10, and 20% missing data.

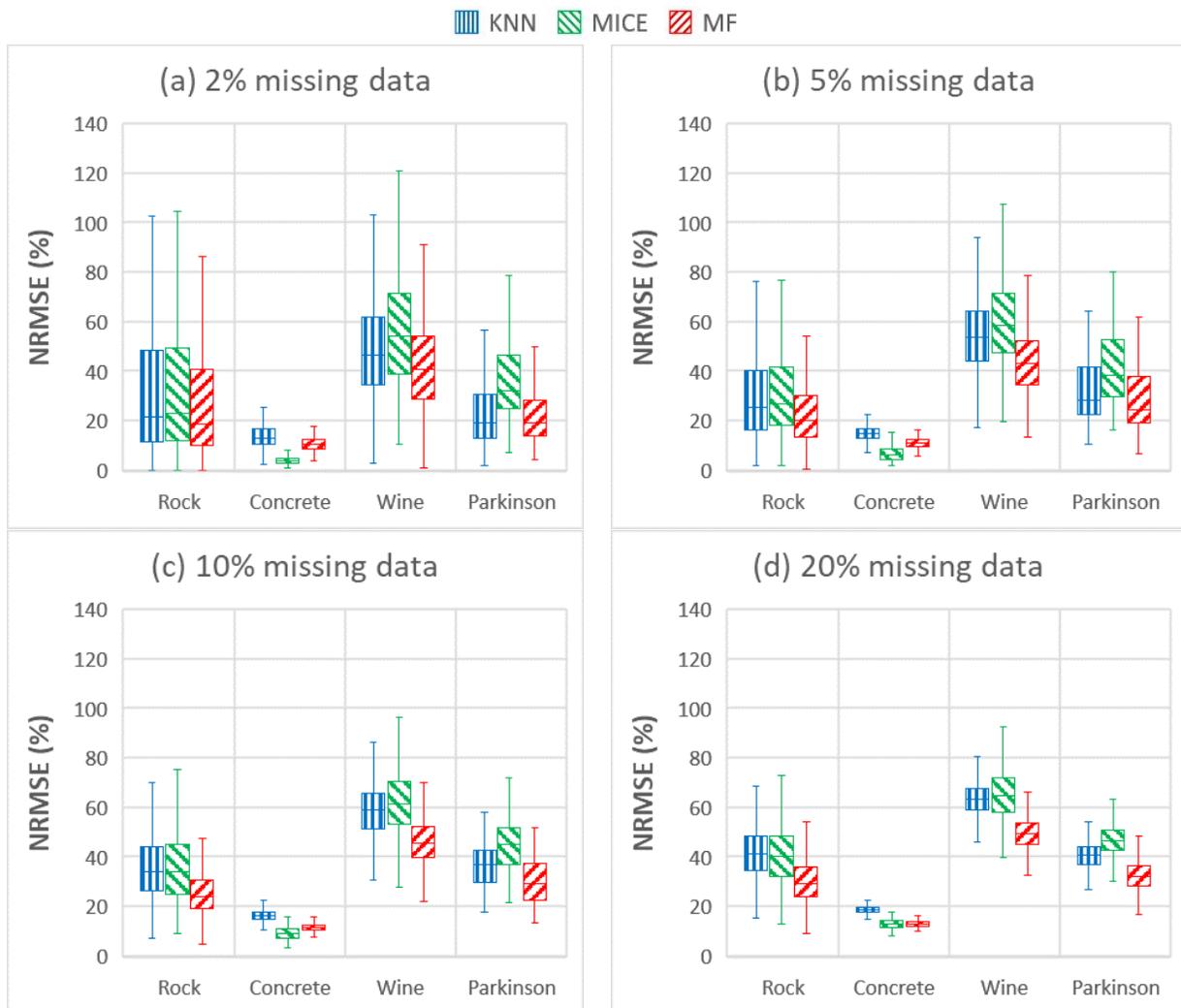

**Figure 3.** Boxplots of NRMSE (%) for the three IMs (KNN, MICE and MF) over the four quantitative datasets ("Rock", "Concrete", "Wine" and "Parkinson") for four different degrees of missing data: (a) 2%, (b) 5%, (c) 10% and (d) 20%.

The four graphs of Figure 3 show the overall trend for imputation errors is to increase on average with increasing percentage of missing data. The only exception is the "Rock" dataset. In this case, even if imputation errors tend to increase from 5 to 20% missing data, they are on average twice as high for 2% missing data than they are for 5% for the three IMs.

As for the qualitative datasets, it appears that the standard deviation of imputation errors decreases with increasing percentage of missing data when observing the length of boxes and whiskers of each graph. On the other hand, even if there is still a trend for the standard deviation of errors to decrease with increasing dataset size, the decrease is less pronounced in comparison to the qualitative cases and the "Concrete" dataset is an exception.

Considering that the four quantitative datasets are arranged in ascending order of dimension, the graphs of Figure 3 highlight the fact that dataset dimension do not seem to affect imputation errors. Indeed, the three IMs performance were the worst on the "Wine" dataset which contains 103 observations and 12 attributes. Imputation errors are then greater than those computed with the "Parkinson" dataset, despite it being the dataset with the highest number of observation and attributes. Similarly, imputation errors computed with the "Rock" dataset are more than twice as high as they are on the "Concrete" dataset which contains double the number of observations and attributes.

As a general comparison, MF outperforms MICE and KNN in almost every case, the exception being the "Concrete" dataset where MICE's imputation errors are on average twice as low as MF's at 2% percentage of missing data. However, the performance gap closes with increasing percentage of missing data until reaching NRMSE around 13% at 20% missing data. Furthermore, KNN is no longer the least performing IM as it outperforms MICE on every dataset but "Concrete".

It is worth noticing that MICE had issues imputing the "Parkinson" dataset because of nearly colinear attributes. This is explained by the parametric nature of MICE and the difficulty these methods have in a situation with highly correlated predictors (Shah et al., 2014). To avoid computational problems, MICE usually removes nearly collinear predictors. However, to take into account all attributes as predictors, the removal was ignored.

2.4.3. Mixed-type datasets

Figure 4 plots imputation errors (PFC and NRMSE) of over 1000 imputations on the four studied quantitative datasets for 2, 5, 10, and 20% missing data (with the exception on the "Musk" dataset on which 500 imputations were performed).

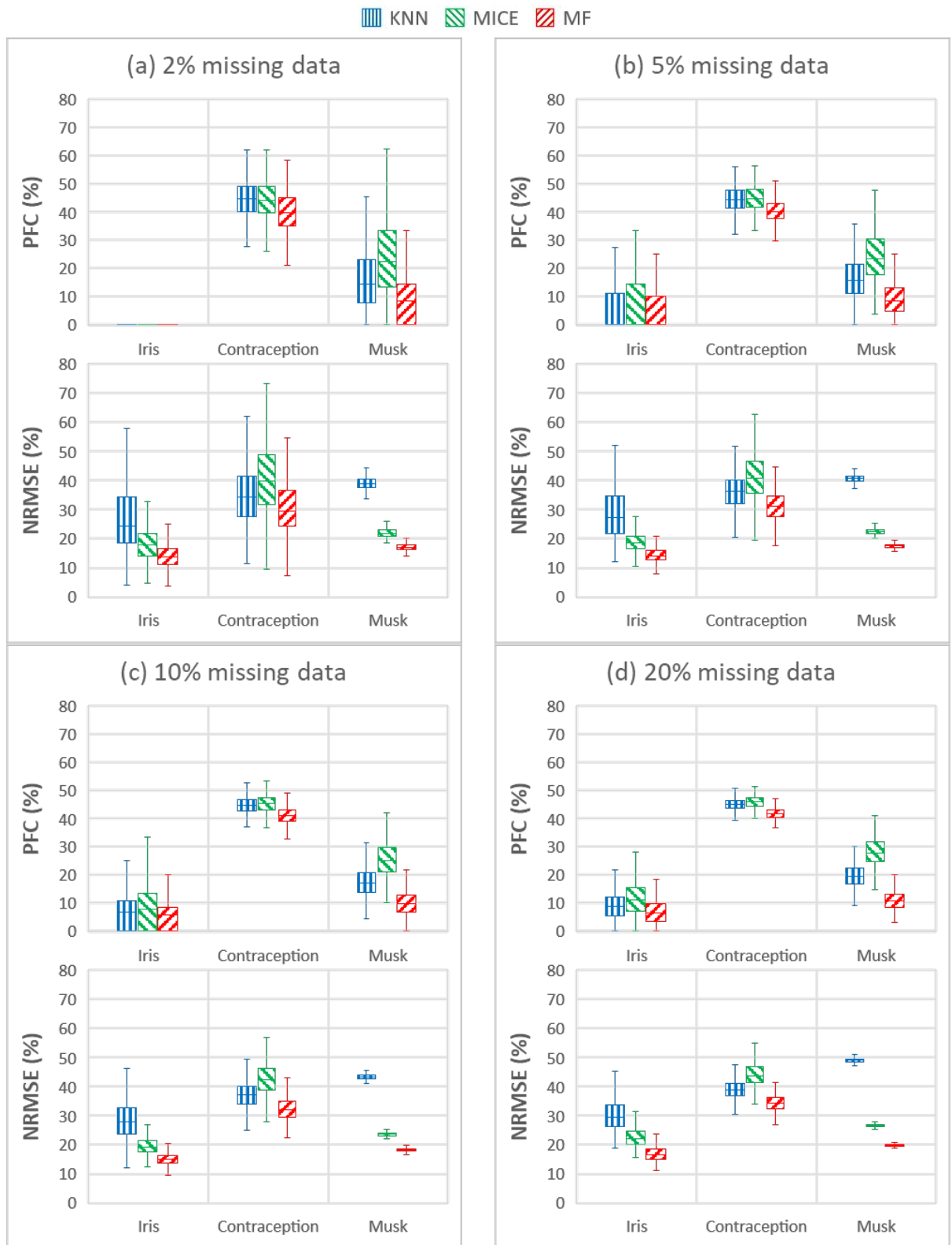

**Figure 4.** Boxplots of PFC (%) NRMSE (%) for the three IMs (KNN, MICE and MF) over the three mixed-type datasets ("Iris", "Contraception" and "Musk") for four different degrees of missing data: (a) 2%, (b) 5%, (c) 10% and (d) 20%.

The observations made about the impact of missing data percentage on imputation errors are the same as for the non mixed-type datasets. There is an overall trend for imputation errors is to increase on average with increasing percentage of missing data. Moreover, the standard deviation of imputation errors still decreases with increasing percentage of missing data when observing the length of boxes and whiskers of each graph.

As for the non mixed-type datasets, when observing the length of boxes and whiskers of each graph it appears that the standard deviation of imputation errors decreases with increasing percentage of missing data. Moreover, although there is still trend for the standard deviation of PFCs to decrease with increasing dataset size, this is not the case or the NRMSEs.

Considering that the three mixed-type datasets are arranged in ascending order of dimension, all graphs of Figure 4 show that dataset size do not seem to affect imputation errors. Even if the NRMSE of KNN tend to increase with increasing dataset size, it is not the case with MICE and MF which are less performant on "Contraception" than they are on "Musk". Furthermore, in terms of PFC, it is also with the "Contraception" dataset that imputation errors were the highest whereas its size is significantly lower than that of "Musk" dataset (respectively 10 and 167 attributes and an equivalent number of records).

A comparison between the respective performances of the three IMs on the graphs of Figure 4 show that MF outperforms MICE and KNN in every case. The gap is however less significant with qualitative attributes. It is for the "Musk" dataset that the performance gap is the highest with an overall 120% decrease in PFC and 83% in NRMSE in regards of the other methods.

2.4.4. Computational efficiency

Table 4 gives the processing times averaged over the 10 studied datasets for 4 different degrees of missing data. All processing time are given in seconds except for the "Musk" dataset for which times are expressed in minutes. Table 5 presents the processing times and rank of KNN, MICE and MF averaged over 4 different degrees of missing data (2, 5, 10 and 20%) for the 10 studied datasets. Both the rank over a specific dataset and the rank within the same dataset type are given in square brackets.

**Table 4.** Processing times (s) of KNN, MICE and MF averaged over the 10 studied datasets for 4 different degrees of missing data

| Missing data percentage | KNN | MICE | MF |
|---|---|---|---|
| 2 | 2.21 | 5.28 | 1.67 |
| 5 | 2.87 | 5.13 | 1.75 |
| 10 | 3.88 | 4.97 | 1.82 |
| 20 | 5.04 | 4.05 | 1.67 |

Different behaviors are illustrated in Table 4 depending on the IM. Processing times for KNN increase with increasing percentage of missing data, whereas those of MICE and MF fluctuate. These different behaviors reflect the differences between each method algorithm. To impute a missing data, KNN must compute the distances between the targeted observation and every other observation in the dataset in order to identify its nearest neighbors. Consequently, the greater the number of observations that are lacking an attribute, the more distances KNN must compute, which costs processing time. On the opposite, the greater the percentage of missing data is, the fewer data there is for MF to test to build its decision trees. Hence not necessarily slowing down model development. As for the fluctuating nature of MICE processing times, it may be explained by its parametric nature. With parametric methods, model development depends on data distribution hypothesis and subsequent parameter calibration, which is not directly affected by the missing data percentage.

**Table 5.** Processing times (s) and rank (1,2 or 3) of KNN, MICE and MF averaged over 4 different degrees of missing data (2, 5, 10 and 20%) for the 10 studied datasets

| Datasets | KNN | | MICE | | MF | | Rank |
|---|---|---|---|---|---|---|---|
| Lanza | 0.03 | [1] | 0.58 | [3] | 0.05 | [2] | [1]: KNN |
| Hayes | 0.08 | [1] | 2.13 | [3] | 0.21 | [2] | [2]: MF |
| Tic-Tac-Toe | 1.02 | [1] | 12.6 | [3] | 2.66 | [2] | [3]: MICE |
| Rock | 0.03 | [1] | 0.15 | [3] | 0.09 | [2] | |
| Concrete | 0.19 | [1] | 0.94 | [2] | 0.94 | [2] | [1]: KNN |
| Wine | 0.17 | [1] | 1.00 | [3] | 0.98 | [2] | [2]: MF |
| Parkinson | 0.53 | [1] | 2.40 | [2] | 4.04 | [3] | [3]: MICE |
| Iris | 0.08 | [1] | 0.53 | [3] | 0.48 | [2] | [1]: KNN |
| Contraception | 0.30 | [1] | 5.92 | [3] | 1.97 | [2] | [2]: MF |
| Musk | 32.6 | [2] | 23.3 | [3] | 5.88 | [1] | [3]: MICE |

Taking into account that datasets are arranged in ascending order of dimension within the same dataset type, Table 5 shows that processing times of each IM tend to increase with increasing size of the datasets. This increase is particularly apparent for the "Musk" dataset for which processing times are expressed in minutes. Indeed, the greater the number of observations, the more distance KNN must compute. Likewise, the greater the number of attributes, the greater the number of distribution parameters MICE must compute. For MF, the processing time necessary to develop an imputation model is dependent on data structure and the associated ease of identifying recurrent data patterns, which is related to dataset size.

Concerning the respective performances of each IM, both tables reveal that MICE is systematically the slowest IM. It is KNN which is the fastest, especially at a low percentage of missing data. This trend might be due to MICE multiple imputation algorithm. Indeed, for every dataset imputed by KNN and MF, MICE imputes 5.

### 2.5. Discussion

Figure 5 presents the general performance comparison of the three IMs. It illustrates the mean reduction of the imputation errors by MF with respect to MICE and KNN as a function of the percentage of missing data. The errors reductions are computed from the average relative gap between the PFC and NRMSE of each IM and averaged over the different data types. Let $d$ denote the relative gap between a value and a reference value (the reference being MF in this case). The relative gap can be defined as:

$$d(x, x_{MF}) = \frac{x - x_{MF}}{x_{MF}} \qquad (6)$$

where $x_{MF}$ is the reference value of MF (PFC or NRMSE) and $x$ the value it is compared to (associated value of KNN and MICE).

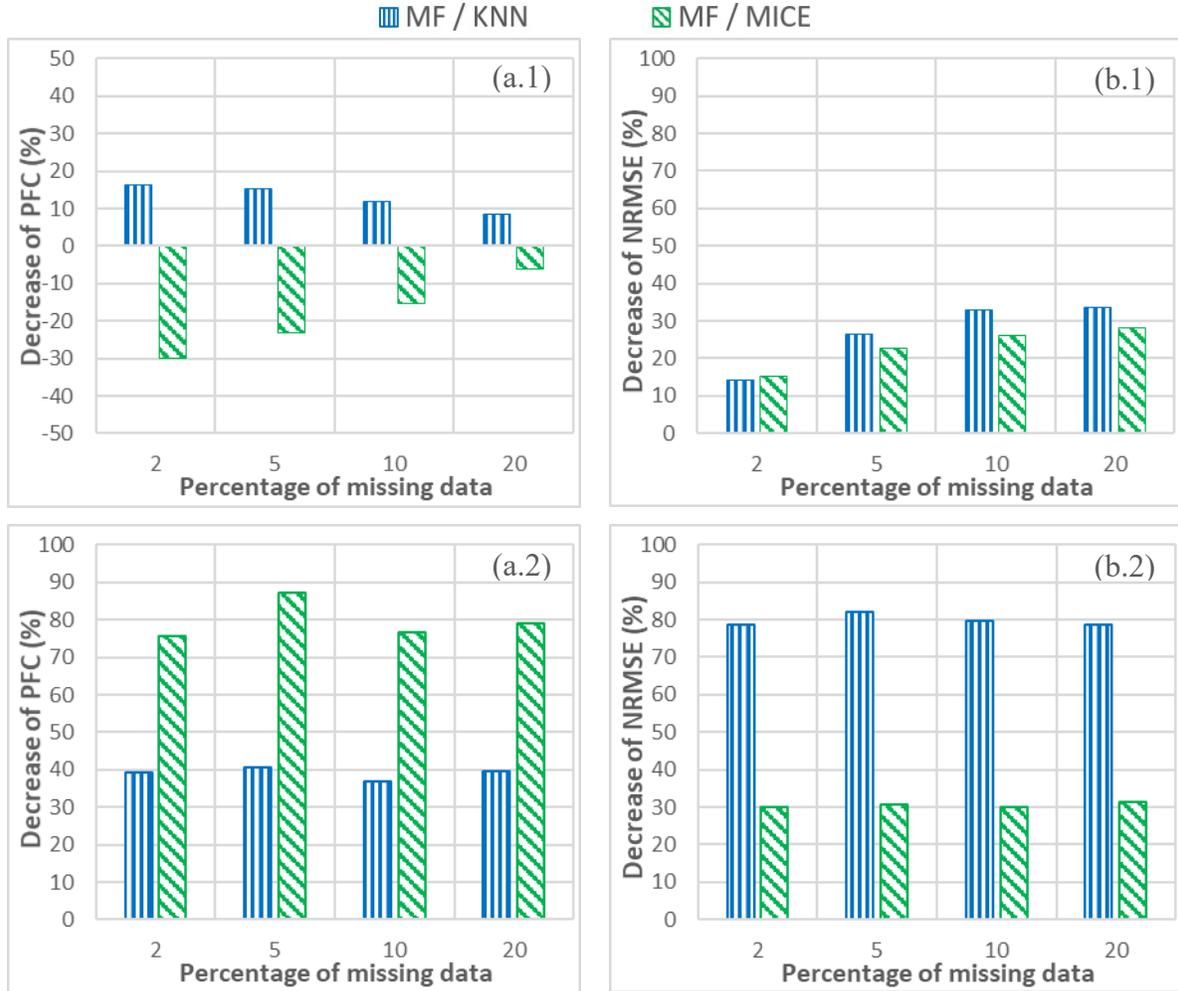

**Figure 5.** Decrease of PFC (%) and NRMSE (%) averaged over (a) qualitative and (b) quantitative data of (1) non mixed-type and (2) mixed-type datasets.

Figure 5 show that MF generally outperforms MICE and KNN on the 10 studied cases. As it can be seen in the (a.1) graph, MF was outperformed only by MICE for exclusively qualitative datasets. Furthermore, this disadvantage compared to MICE tends to decrease with increasing percentage of missing data. Figure 5 also reveals that the performance gap between MF and the two other IMs is more pronounced on the mixed-type datasets, with errors reductions systematically superior to 30%, even for errors associated with qualitative attributes (PFC). Based on these observations and given the heterogeneous nature of environmental databases (with both qualitative and quantitative entries), it seems appropriate to consider MF when applying IMs to environmental contexts.

In order to highlight a potential relationship between data structure and the specific performance of MF, interactions between attributes were characterized for the quantitative datasets. This characterization also permitted to investigate a potential link between data structure and the variability of imputation errors within the same dataset type. Indeed, imputation errors of the three IMs showed significant variability not linked to dataset size, for the quantitative datasets. Table 6 presents the indices describing the structural characteristics of the qualitative datasets, the associated imputation errors averaged on every simulations and percentage of missing data and the mean decrease in errors by MF. The three used indices are: the absolute mean correlation – $\rho_{abs}$, the absolute standard deviation – $sd_{abs}$ and the absolute skewness index – $skew_{abs}$. They give an insight on correlation intensity, unbalancing between correlation and on the shape of the unbalancing.

**Table 6.** Structural characterization of quantitative datasets juxtaposed with mean NRMSEs and mean decreases of NRMSEs by MF

| Datasets (rows × columns) | $\rho_{abs}$ | $sd_{abs}$ | $skew_{abs}$ | Mean NRMSE | Mean decrease of NRMSE MF/KNN | Mean decrease of NRMSE MF/MICE |
|---|---|---|---|---|---|---|
| Rock (48 × 4) | 0.52 | 0.22 | -0.07 | 39 % | 32 % | 32 % |
| Concrete (103 × 10) | 0.25 | 0.17 | 1.52 | 12 % | 36 % | -29 % |
| Wine (122 × 12) | 0.21 | 0.18 | 1.18 | 55 % | 23 % | 39 % |
| Parkinson (195 × 22) | 0.50 | 0.30 | 0.01 | 34 % | 15 % | 50 % |

The structural characteristics presented in Table 6 show that MF's performance compared to KNN and MICE does not seem affected by the intensity of correlations in the imputed datasets. In regard to KNN, errors reductions are the highest on "Rock" and "Concrete" which are respectively moderately ($\rho_{abs}$ around 0.5) and weakly correlated ($\rho_{abs}$ around 0.2). Likewise, MF's reductions of MICE's imputation errors are the greatest on the "Wine" and "Parkinson" datasets, whereas one is weakly correlated and the other moderately correlated (Wassertheil and Cohen, 1970). Furthermore, no observation has been made over the unbalancing amongst the correlations of each datasets because their parameters are of the same order of magnitude. In the light of these results, the studied structural criteria is not sufficient to make inferences on MF's performance in comparison to MICE and KNN. The lack of a relationship between data structure and imputation accuracy may be due to the fact that "Parkinson" and "Rock" structures are not complex enough to make a significative difference over MF specific performance. No relationship between the indices and the general performance of MF, MICE and KNN has been identified. Indeed, the best and worst imputation errors for the three IMs were obtained on the "Concrete" and "Wine" datasets (on average 12 and 55% respectively), though they have nearly similar indices. The results of earlier studies (Solaro *et al.*, 2015 ; Solaro *et al.*, 2017) suggested a general decrease in imputation errors as the value of correlation indices increases, especially with skew data. This divergence indicates that imputation accuracy is not determined only by the studied data structure descriptive criteria. The same observation can be made about data size, which did not seem to affect IM performance in comparison to data type and percentage of missing data. Additionally, estimating structural aspects of incomplete datasets is not recommended in practice as it could produce biased results (Leite and Beretvas, 2017).

As a general remark, we can state that, despite its decline in performance with exclusively qualitative datasets, MF is a robust IM in comparison to KNN and MICE, which are among the most used IM to this day. MF showed the best overall performances over a variety of cases which seem to indicate that it is the method suitable to be used when dealing with missing data in environment. However, because the issues tackled in environment are diverse, so are environmental databases. Hence, it is not possible to assert MF will be the best IM in every case, especially given the multitude of parameters that could impact imputation accuracy. Furthermore, for each imputation, MF runs an OOB estimate for the imputation error which may be useful in practice.

### 3. Case study: Quebec wastewater treatment plants performance monitoring

The results of the comparative study suggested that MF is the most robust of the three studied IMs. It was thus applied to the case study of Quebec wastewater treatment plants performance monitoring. The case study imputation is presented in this section into four parts. Section 3.1 describes the imputed database, Section 3.2 the evaluation criteria used to assess the imputation quality. The results of the imputations are given in Section 2.3 and discussed in Section 2.4.

### 3.1. Quebec wastewater treatment plants performance database

The studied dataset made available through Canadian government sources characterizes the effluents of 814 wastewater treatment plants for each day of the year 2013. The effluents are described under 11 operational parameters. A preliminary analysis revealed that some of the treatment plants did not possess any data for the performance parameters of that year. The data retrieval of the latter left 789 treatment plants. For calculating the performance requirements established by the ministry, an annual average has been applied for each parameter and for each treatment plant. Table 7 presents the 11 operational parameters (from most to least complete) contained in the annual database, as well as their unit, their abbreviation, and their missing data percentage. Additionally, the last column of Table 7 gives the missing data percentage for the dataset formed by the parameter in a given row and all the ones in the previous rows. For example, if we consider a databased formed of parameters TRT, COD, BOD, SS and pH, the missing data percentage is 0.25.

Table 7. Description of the 11 operational parameters

| Parameter | Unit | Abbreviation | Missing data percentage (parameter) | Cumulative missing data percentage (dataset) |
|---|---|---|---|---|
| Treatment type | 9 classes | TRT | 0.00 | 0.00 |
| Chemical oxygen demand concentration | $mg \cdot L^{-1}$ | COD | 0.00 | 0.00 |
| Biochemical oxygen demand concentration for 5 days | $mg \cdot L^{-1}$ | $BOD_5$ | 0.13 | 0.04 |
| Suspended solids concentration | $mg \cdot L^{-1}$ | SS | 0.13 | 0.06 |
| Ammoniacal nitrogen concentration | $mg \cdot L^{-1}$ | $NH_4$ | 0.25 | 0.10 |
| pH | without unit | pH | 1.01 | 0.25 |
| Logarithm of the fecal coliforms concentration | $CFU \cdot 100mL^{-1}$ | log.FC | 4.69 | 0.89 |
| Total phosphorus | $mg \cdot L^{-1}$ | Ptot | 23.8 | 3.75 |
| Measured flow rate | $m^3 \cdot d^{-1}$ | Q.MSR | 96.7 | 14.1 |
| Calcium carbonate concentration | $mg \cdot L^{-1}$ | $CaCO_3$ | 98.4 | 22.5 |
| Total kjeldahl nitrogen | $mg \cdot L^{-1}$ | Ntk | 99.2 | 29.5 |

A preliminary analysis of the averaged dataset highlighted some parameters with a percentage of missing data higher than 96%. Establishing an arbitrary missing data percentage threshold of 90% resulted in the removal of 3 attributes, retaining 8. The retained parameters are the following: TRT, COD, $BOD_5$, SS, $NH_4$, pH, log.FC and Ptot. Table 8 presents the general characteristics of the original and pretreated datasets.

Table 8. Characteristics of the original and pretreated datasets

| Characteristics | Original dataset | Pretreated dataset |
|---|---|---|
| Number of records | 789 | 789 |
| Number of parameters | 11 | 8 |
| Missing data percentage | 29.5 | 3.75 |

### 3.2. Evaluation criteria

When one does not possess a complete dataset to compare the imputed data to the real ones, the accuracy of an imputation can be estimated by means of an OOB estimates such as the one missForest computes, the OOB error. This estimate removes the need for the usual cross-validation error estimate (Refaeilzadeh et al., 2009), which requires the user to split up the data into two separate sets, a training and a test set. In this type of procedure, because the test set is used to evaluate model accuracy, only the information contained in the training set is used to build the model. Furthermore, whereas cross-validation techniques may require significant computing efforts, OOB errors can be computed during model building thanks to the bagging of predictors. The results of previous works (Breiman, 1996a; Stekhoven and Bühlmann, 2012) brought forward the noteworthy accuracy of OOB errors, generally differing from the actual errors only by a few percent. MF performance applied to the wastewater treatment plant dataset has been assessed with the OOB error.

### 3.3. Results

The preprocessed dataset of Quebec wastewater treatment plants performance monitoring was imputed with MF, considering processing times (s) and estimated imputation errors (%) given by the OOB error. Considering the random factor of the MF algorithm, the imputation was repeated 1000 times. Figure 6 shows the boxplots of OOB errors and processing times over 1000 imputations.

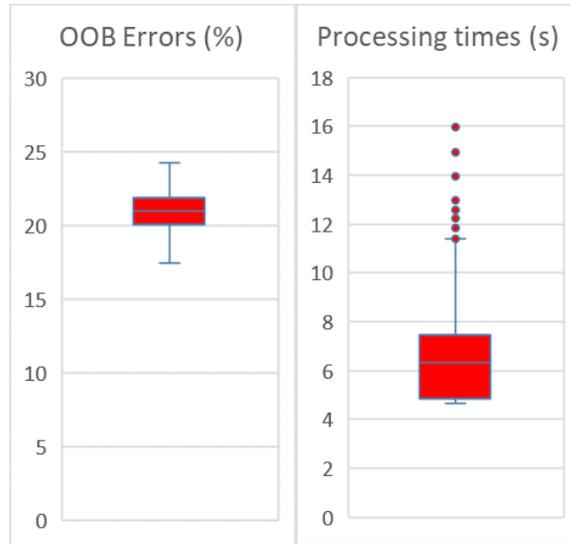

**Figure 6.** Boxplots of OOB errors (%) and processing times (s) for the 1000 imputations performed by MF on the pretreated Quebec wastewater treatment plants performance dataset.

Figure 6 shows that OOB errors vary between 17 and 24% with a standard deviation around 1.4. Estimated imputation errors around 21% is a satisfactory result when compared to the imputation errors that were obtained during the comparative study. In fact, amongst the ten studied datasets, only three cases yielded better results, namely Concrete, Iris and Musk. Processing times are systematically inferior to 20 seconds, the time necessary for imputation is therefore not a constraint in this case.

Amongst the 8 imputed parameters, Ptot is the parameter with the highest percentage of missing data. Removing this parameter reduces the imputed dataset missing data percentage from 3.75 to 0.89%. It is thus relevant to study the impact of the Ptot parameter on OOB errors. Figure 7 presents the boxplots of OOB errors and processing times over 1000 imputations on the Quebec wastewater treatment plants performance dataset without the Ptot parameter.

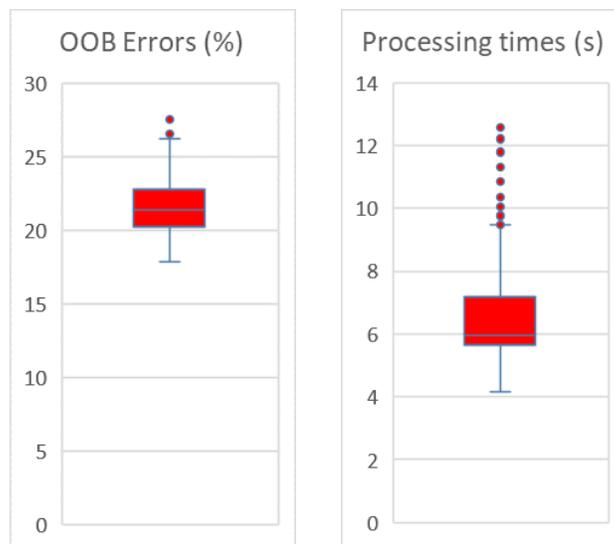

**Figure 7.** Boxplots of OOB errors (%) and processing times (s) for the 1000 imputations performed by MF on the pretreated Quebec wastewater treatment plants performance dataset without the Ptot parameter.

Figure 7 shows that OOB errors are slightly higher (0.5% on average) than for the imputations including Ptot. Similarly, the standard deviation of the 1000 OOB errors increased from 1.4 to 1.9. Contrary to what was suggested by the comparative study results, this rise in OOB errors indicates that missing data percentage and imputation accuracy are not directly related. This may be explained by the fact that, with MF, model accuracy is less dependent on missing data percentage than it is on data structure and the associated ease of identifying recurrent data patterns which will be used to make inference. Consequently, even if Ptot was the one parameter with the smallest number of entries, the observed entries might have significantly helped for the identification of generalizable data patterns.

### 3.4. Discussion

The specificities of MF algorithm made possible the imputation of the Quebec wastewater treatment plants dataset and produced satisfying imputation errors. Additionally, the imputations performed on different versions of the case study dataset revealed that the missing data percentage of an imputed dataset does not define the accuracy of the imputations. From an operational point of view, the OOB errors provides an additional asset, efficiently computing an error estimate. The new complete dataset now allows for an exhaustive analysis of the wastewater treatment plants performance and thus gives a better understanding of how the different types of wastewater treatments applied in Quebec impact water quality. Furthermore, what has been done for the year 2013 can be done for each year until 2001, hence granting an interannual scope to the analysis. It is worth noticing that depending on the objectives of the decision makers, this study could be transposed to different time scales and/or with additional parameters. It would, for instance, be possible to illustrate the influence of meteorological conditions on the monitored parameters.

### 4. Conclusion

In this study, we investigated and compared the applicability of three established IMs (KNN, MICE and MF) to the issue of missing environmental data. As a means of assessing IM robustness, a computational study was performed using 10 datasets differing in terms of size, attributes type, structural features, and sources, considering imputation errors and processing times. Following the performance comparison, a case study of Quebec wastewater treatment plants performance monitoring was carried out using the overall most performing IM.

The results of the comparative study show that except for exclusively qualitative datasets, MF generally outperforms KNN and MICE. This is especially true with mixed-type datasets, for which MF reduced imputation errors up to 150%. Although it was not the fastest IM, MF showcased competitive computational efficiency. This first experimental phase did not highlight the defining dataset characteristics which determine imputation accuracy, as it was not dimension nor structural features. However, in regards of its performances facing mixed-type datasets, the case study was performed using the MF method. After pretreating the original database, MF imputed the Quebec wastewater treatment plants missing data with satisfying accuracy according to the OOB error.

In the lights of the above findings, we can say that MF robustness in comparison to KNN and MICE makes it the most suited IM for the diversity of datasets dealt with in environmental applications. Its algorithm allows for the reconstruction of any type of data, thus giving the opportunity to improve data analysis and subsequent decision making. Additionally, it provides an OOB estimate which gives a way to evaluate the accuracy of an imputation and is an easy to implement method.

Finally, as future work, we contemplate the following: (i) study of the performance of IMs facing big data; (ii) study of the impact of other missingness types on IM performance.

**Acknowledgments.** The authors acknowledge the R development core team and UCI machine learning repository for making data available to the user community. Authors are also grateful to the MELCC for providing the Quebec wastewater treatment plants performance monitoring database. Finally, the authors acknowledge the support of the Natural Sciences and Engineering Council of Canada (NSERC), [funding reference number RGPIN-2014-06154].